\documentclass[aps,pre,twocolumn,
superscriptaddress,groupedaddress,showpacs]{revtex4}
\usepackage{graphicx,epsf}
\usepackage{xcolor}
\usepackage{psfrag}
\begin{document}
%----------------------------------------------------------------------------
%                              TITLE
%----------------------------------------------------------------------------
\title{ Ring polymers in crowded environment: conformational properties}
%----------------------------------------------------------------------------
%                              AUTHORS revtex4-style
%----------------------------------------------------------------------------
\author{K. Haydukivska}
\affiliation{Institute for Condensed
Matter Physics of the National Academy of Sciences of Ukraine,\\
79011 Lviv, Ukraine}
\author{V. Blavatska}
\email[]{E-mail:  viktoria@icmp.lviv.ua}
\affiliation{Institute for Condensed
Matter Physics of the National Academy of Sciences of Ukraine,\\
79011 Lviv, Ukraine}

%----------------------------------------------------------------------------
%                             ABSTRACT
%----------------------------------------------------------------------------
\begin{abstract}
 We analyze the universal size  characteristics of flexible ring polymers
 in solutions in presence of structural obstacles (impurities) in $d$ dimensions. One encounters such situations when considering polymers in gels, colloidal solutions, intra-
and extracellular environments. A special case of extended impurities correlated on large distances $r$ according to a
 power law $\sim r^{-a}$ is considered. Applying the direct polymer renormalization scheme, we evaluate the
 estimates for averaged gyration radius $\langle R_{g\,{\rm ring}} \rangle$ and spanning radius   $\langle R_{1/2\,{\rm ring}} \rangle$ of typical ring polymer conformation up to the first order of double $\varepsilon=4-d$, $\delta=4-a$ expansion.  Our results quantitatively reveal an extent of the effective size and anisotropy of closed ring macromolecules  in disordered environment. In particular, the size ratio
  of ring and open (linear) polymers of the same molecular weight grows when increasing the strength of disorder according to $\langle R^2_{g\,{\rm ring}} \rangle / \langle R^2_{g\,{\rm chain}} \rangle =\frac{1}{2} \left(1+\frac{13}{48}\delta \right)$.
\end{abstract}
\date{\today}
%\maketitle%

\maketitle
\section{Introduction}

Long flexible macromolecules in nature often appear in the form of closed loops (rings). One can find such polymers inside the living cells of bacteria \cite{Fiers62}
or sometimes higher eukaryotes \cite{Zhou03}, where DNA occurs in a ring shape.
Loop formation is an important feature of chromatin organization \cite{Fraser06,Simonis06}, playing a vital role in transcriptional regularization of genes and
DNA compactification in the nucleus.
On the other hand, many synthetic polymers form circular structures
during polymerization and polycondensation \cite{Brown65,Geiser80,Roovers83}. Such a wide range of physical realizations,
where the macromolecules of closed ring type can be found, make them  a subject of intensive experimental and analytical
studies \cite{Bishop86,Diel89,Jagodzinski92,Obukhov94,Muller99,Deutsch99,Miyuki01,Calabrese01,Alim07,Bohr10,Sakaue11,Jung12,Ross11}.

Statistics of long flexible polymers in good solvents is known to be characterized by a set of universal properties, independent on details of microscopic chemical structure of macromolecules \cite{deGennes,desCloiseaux}. In particular, the averaged radius of gyration $\langle R_{g\,{\rm chain}}^2\rangle$ and the end-to-end distance $\langle R_{e\,{\rm chain}}^2\rangle$ of  linear polymer chains obey the scaling law:
\begin{equation}\label{scalingR}
  \langle R_{g\,{\rm chain}}^2 \rangle \sim \langle R_{e\,{\rm chain}}^2 \rangle
\sim N^{2\nu}, \label{scalR}
\end{equation}
here $\langle (\ldots) \rangle$ means averaging over an ensemble of possible conformations of macromolecule, $N$ is the molecular weight (number of repeating units -- monomers) and $\nu$ is the universal critical exponent, that only depend on the
space dimension $d$.
 E.g., in $d=3$ the refined field-theory renormalization group studies give $\nu =0.5882\pm 0.0011$ \cite{Guida98}, whereas for the case of space dimension above the upper
critical one $d_{\rm up}{=}4$ one has an ideal Gaussian polymer chain without excluded volume effect with $\nu (d\geq4){=}1/2$  {(note that corrections to scaling are logarithmic at critical dimension so that
$\langle R_{g\,{\rm chain}}^2\rangle(d=4)\sim N (\log(N))^{1/4}$)}. The radius of gyration  $\langle R_{g\,{\rm ring}}^2\rangle$ of closed polymer rings
obeys the scaling law (\ref{scalingR}) with exactly the same critical exponent $\nu$  \cite{Prentis82,Privman}. As the convenient parameter to compare the  size measures of linear and ring polymers of the same molecular weight $N$,  one can consider the ratio:
 \begin{equation} \label{ratioR}
 g\equiv \frac{\langle R_{g\,{\rm ring}}^2\rangle}{\langle R_{g\,{\rm chain}}^2\rangle},
 \end{equation}
 which is universal
$N$-independent quantity. It was found that in the idealized  Gaussian case  $g=1/2$  \cite{Zimm49},
whereas presence of excluded volume effect leads to an increase of this value \cite{Baumgaertner81,Prentis82,Prentis84}.
Note that for the closed circular polymers, the spanning radius  $ R_{1/2\,{\rm ring}} $ is of interest instead of the usual end-to-end distance, and the ratio
\begin{equation} \label{ratioRpiv}
 p\equiv \frac{\langle R_{1/2\,{\rm ring}}^2\rangle}{\langle R_{g\,{\rm ring}}^2\rangle}
 \end{equation}
is another universal relation, which characterizes the spatial distribution of monomers within the macromolecule.
  {
As established by de Gennes \cite{deGennes}, the universal statistical properties of
infinitely long flexible polymers are perfectly captured by the
 $m$-component spin vector model at its critical point in the formal limit $m \to 0$.
 In particular, the polymer size exponent
$\nu$ as given by (\ref{scalR}) is related to the correlation length critical index
of the $m=0$ model, whereas the universal size ratios
(\ref{ratioR}) and (\ref{ratioRpiv}) can be computed in terms of the critical amplitudes ratios of this model (see e. g. \cite{Aronovitz86}). }

An important question in polymer physics is how the universal conformational properties of macromolecules are modified in presence of structural
obstacles (impurities) in the system.
One can encounter such situation when considering polymers in gels, colloidal solutions \cite{Pusey86},  intra-
and extracellular environments \cite{Kumarrev,cel1,cel2}.
Biological cells can be described as  disordered (crowded) environment due to the presence of a large amount of various biochemical species \cite{Minton01}.
It is established, that presence of structural defects strongly effect the protein folding and aggregation \cite{Horwich,Winzor06,Kumar,Echeverria10}.

%It was shown analytically \cite{Kim87} and confirmed in numerical simulations \cite{Kremer,Lee88,Woo91}, that %structural disorder in the form
%of randomly distributed point-like defects of low concentration does not alter the universality class of polymer %macromolecules.
%Only when concentration of such impurities reaches the so-called percolation threshold value, the presence of disorder causes non-trivial effect on
%statistical properties of polymer macromolecules  \cite{Kremer,Grassberger93,Ordemann02,Janssen07}.
The structural impurities in environment often cannot be treated like point-like defects: they can be comparable in size with polymer chain or even penetrate throughout the system. The density fluctuations of disorder may lead a considerable spatial inhomogeneity and create pore spaces,
which are often of fractal structure \cite{Dullen79}.
 { These peculiarities are perfectly captured within the so-called percolation model \cite{Stauffer}, which already
 serves as a paradigm in studies of disordered systems. At critical concentration $p_c$ of structural obstacles,
  an incipient percolation cluster of fractal structure can be found in the system.
  Numerous analytical and numerical studies \cite{Kremer,Woo91,Grassberger93,Rintoul94,Ordemann02,Janssen07,Blavatska10a} indicate the considerable extension of effective polymer size
 (in particular an increase of scaling exponent $\nu $ in (\ref{scalingR}) and increase of elongation and anisotropy of typical polymer conformations caused by complex structure of underlying percolation cluster.}

Another special type of disorder which display correlations in mesoscopic scale can be described within the frames of a model with long-range correlated
quenched defects, proposed in Ref. \cite{Weinrib83} in the context of magnetic phase transitions.
 Here, the defects are assumed to be correlated on large distances $r$
according to a power law with a pair correlation function $
g(r)\sim r^{-a} $ \cite{Weinrib83}.
For $a<d$, such a correlation function describes defects extended in space, which form complex structures of
(fractal) dimension $d_f=d-a$, such that
$a=d-2$ $(d-1)$ correspond respectively to the impurities in form of lines (planes), randomly distributed in  space, whereas non-integer values of $a$ refer to defects of fractal structure.
 { The influence of long-range-correlated
disorder on the critical properties of $m$-component spin model has been analyzed in Refs. \cite{Weinrib83,Prudnikov}
within the refined field-theoretical approach. Here, the variable $a$ was argued to be a global parameter along with the space dimension $d$ and the number of components $m$ of the order parameter, and thus the presence of
long-range-correlated disorder leads to a new universality class for these magnetic systems.
In particular, the correlation length critical exponents in this case are larger than corresponding values in absence of disorder, and increase with decreasing the parameter $a$.
 The effect of long-range-correlated disorder on the scaling properties of linear polymer chains was established by analyzing the critical properties of $m=0$ model in Ref. \cite{Blavatska01a}. Presence of disorder in the form
  of extended structural defects causes the swelling of polymer coil (\ref{scalingR}) with larger value
of scaling exponent $\nu$, and thus leads to an elongation of polymer conformation. Further studies reveal an increase of the effective polymer size and shape anisotropy of polymers in long-range correlated disorder \cite{Blavatska10}}.  Moving from linear polymer chains to more complicated structures like star-branched polymers in environment with long-range correlated disorder,
one finds  the whole spectrum of universal exponents
in a new universality class \cite{Blavatska06}. In this concern, it is worthwhile to study the influence of extended defects on the statistical properties of
polymers of circular structure, which have not been considered so far.

In this paper we analyze the statistical properties of ring polymers in environment with long-range correlated disorder analytically, applying the direct polymer renormalization scheme. The special attention is paid to the universal size  characteristics such as the size ratios (\ref{ratioR}) and (\ref{ratioRpiv}).

The layout of the paper is as follows. In the next section, we introduce the continuous model of flexible circular polymer in disordered environment. In section III the method of direct polymer renormalization is shortly described. The results
for universal conformational properties such as the size ratios are evaluated in Section IV. We end up by giving the conclusions in Section V.

\section{The model}
We consider flexible ring polymers in solutions in presence of long-range correlated disorder.
Within the Edwards continuous chain model \cite{Edwards}, the linear polymer chain is considered as a path of length $S$, parameterized by $\vec{r}(s)$, where $s$ is varying from $0$ to $S$.
The partition function of closed polymer ring is given by \cite{Duplantier94}:
\begin{eqnarray}
&&Z=\int\!\! D\vec{r}\, \delta (\vec{r}(S)-\vec{r}(0)) \exp \left[ -\frac{1}{2}
\int^{S}_{0}\!\!\left(\frac{{\rm d}\vec{r}(s)}{{\rm d}s}\right)^{2}\!{\rm d}s \right.-
\nonumber\\
&&-\frac{b_0}{2}\int^{S}_{0}\!\!{\rm d}s'\int^{S}_{0}\!\!{\rm d} s''\,\delta(\vec{r}(s')-\vec{r}(s''))+\nonumber\\
&&\left.+\int^{S}_{0}V(\vec{r}(s))\,{\rm d} s\right]. \label{model-con}
\end{eqnarray}
Here, $\int D\vec{r}$ is functional path integrations, the $\delta$-function describes the fact that the path is closed, the first term in the exponent governs the behavior of Gaussian polymer, the second term describes short-range repulsion between monomers due to excluded volume effect governed by coupling constant $b_0$ and the last one
arises due to the presence of disorder in the system and contains a random potential $V(\vec{r}(s))$.
Let us denote by ${\overline{(\ldots)}}$ the average over different realizations of disorder and assume \cite{Weinrib83}:
\begin{equation} {\overline{ V(\vec{r}(s))V(\vec{r}(s'))}} = w_0 |\vec{r}(s')-\vec{r}(s'')|^{-a}. \label{avv0}
\end{equation}

Studying the problems connected with randomness (disorder) in the system, one usually faces two types of ensemble
averaging. In so-called annealed case \cite{Brout59}, the impurity variables are a part of the disordered system phase
space, which amounts averaging the partition sum of a system over the random variables. In the quenched
case \cite{Emery75}, the free energy (the logarithm of the partition sum)
should be  averaged over an ensemble of realizations of disorder, which usually implies the replica formalism.
 In general, the critical behavior of systems with
quenched and annealed disorder is quite different.
 However, when studying the universal conformational properties of  long flexible macromolecules,
  this distinction is negligible \cite{Blavatska13} and one can use the annealed averaging, which is technically simpler.
Performing the averaging of the partition function (\ref{model-con}) over different realizations of disorder, taking into account
up to the second moment of cumulant expansion and recalling (\ref{avv0}) we obtain:
\begin{equation}
{\overline {{Z}}} =\int D\vec{r}\, \delta (\vec{r}(S)-\vec{r}(0))\,{\rm e} ^{-H}
\end{equation}
with an effective Hamiltonian:
\begin{eqnarray}
&&H=\frac{1}{2}\int_0^{S}{\rm d}s\!\!
\left(\frac{{\rm d} {\vec {r}}(s)}{{\rm d} s}\right)^2 + \nonumber \\
&&+ \frac{b_0}{2}
\int_0^{S}\!\!{\rm d}s'\int_0^{S}\!\!{\rm d}s{''}\,\delta({{\vec{r}}}(s')-{\vec{r}}(s{''})){-}\nonumber\\
 &&-\frac{w_0}{2}\int_0^{S}\!\!{\rm d}s'\int_0^{S}\!\!{\rm d}s{''}\, |\vec{r}(s')-\vec{r}(s'')|^{-a}.
\label{Hdis}
\end{eqnarray}
Note that the last term in (\ref{Hdis}) describes an effective attractive interaction
between monomers arising due to the presence of extended obstacles in environment, governed by coupling constant  $w_0$.

Performing dimensional analysis for the  terms in (\ref{Hdis}) one finds the dimensions of the couplings in terms of dimension of contour length $S$: $[b_0]=[S]^{{\rm d}_{b_0}}$, $[w_0]=[S]^{{\rm d}_{w_0}}$
with ${\rm d}_{b_0}=(4-d)/2$, ${\rm d}_{w_0}=(4-a)/2$. The ``upper critical" values of the space dimension ($d_c=4$) and the correlation parameter ($a_c=4$), at which the couplings are dimensionless, play an important role in the renormalization scheme, as outlined below.

\section{The method}

To analyze the universal statistical properties of model (\ref{Hdis}), we evaluate the direct
renormalization method, as developed by des Cloizeaux \cite{desCloiseaux}.

In the asymptotic limit of an infinite linear measure of the continuous polymer curve,
one encounters the divergences of  observables of interest. All these divergences can be eliminated by
introducing corresponding renormalization factors, directly associated with physical quantities.
Subsequently, they  attain finite values when evaluated at the stable fixed point (FP) of the renormalization group transformation.
Note that the FP coordinates are universal, so that properties  of a linear polymer chain and that of a closed polymer ring are
 governed by the same unique FP. Therefore, to evaluate the FP coordinates in the following analysis
we restrict ourselves to the simpler case of a single chain polymers. To define
the coupling constant renormalization,
one considers the contributions
${\cal Z}_{\lambda_0}(S,S)$  into  the partition function of two interacting chain polymers, having dimensions
 ${\cal Z}_{\lambda_0}(S,S)\sim [S]^{2+d_{\lambda_0}}$ (in our case, by $ \lambda_0 $ we mean couplings $b_0$ and $w_0$).
The renormalized coupling constants $\lambda_R$ are thus defined by:
\begin{eqnarray}
&&\lambda_R(\{ \lambda_0\})=-[Z(S)(\{\lambda_0 \})]^{-2}{Z}_{\lambda_0}(S,S)\times\nonumber\\
&&\times[2\pi\chi_0(\{\lambda_0 \}S ]^{-(2-{\rm d_{\lambda_0}}) }. \label{ures}
\end{eqnarray}
Here, $Z(S)(\{\lambda_0 \})$ is the partition function of a polymer,
$\chi_0(\{\lambda_0 \}$ is the so-called swelling factor, given by: $\chi_0(\{\lambda_0 \}=\langle R_e^2 \rangle /S$.

In the limit of infinite linear size of the macromolecules  the renormalized theory remains finite, such that:
\begin{equation}
\lim_{S\to\infty} \lambda_{R}(\{ \lambda_0\})=\lambda_R^*.
\end{equation}
When couplings constants are dimensionless (which happens at corresponding ${\rm d_{\lambda_0}}= 0$),
 the macromolecules  behave like Gaussian chains without any interactions between monomers.
 Thus, the concept of expansion in small
deviations from the upper critical dimensions of the coupling constants naturally arises.

 The flows of the renormalized coupling constants are governed by functions $\beta_{\lambda_R}$:
\begin{equation}
\beta_{\lambda_R}=2S\frac{\partial \lambda_R(\{ \lambda_0\})}{\partial S}.\label{fip}
\end{equation}
The fixed points of the renormalization group transformations, which define the asymptotical values of
universal conformational properties, are given
 by the common zeros of the $\beta$-functions.

\section{Results}

In the present work we  analyze the  conformational characteristics of long flexible ring polymers in the environment with long-range correlated disorder. We are interested in size ratios (\ref{ratioR}) and (\ref{ratioRpiv})
which are known to be universal.

Within the frames of the continuous polymer model (\ref{model-con}),
 the  averaged radius of gyration $R_g$, the end-to-end distance $R_e$ of an open linear chain and the spanning radius $R_{1/2}$ of circular polymer
 are defined as:
\begin{eqnarray}
&&\langle R_g^2 \rangle = \frac{1}{2S^2} \int_0^S \!\! ds_1\int_0^{S} \!\! ds_2
\left\langle(\vec{r}(s_2)-\vec{r}(s_1))^2 \right\rangle,  \label{rg}\\
&&\langle R_e^2 \rangle = \left\langle
(\vec{r}(S)-\vec{r}(0))^2 \right\rangle,  \label{re} \\
&&\langle R_{1/2}^2 \rangle = \left\langle
(\vec{r}(S/2)-\vec{r}(0))^2 \right\rangle.  \label{1/2}
\end{eqnarray}
Here and below, $\langle \ldots \rangle$ denotes averaging with an effective Hamiltonian (\ref{Hdis}) according to:
\begin{equation}
\langle (\ldots) \rangle = \frac{\int D\vec{r}\, \delta (\vec{r}(S)-\vec{r}(0))\,(\ldots)\,{\rm e} ^{-H}}{{ {{Z(S)}}} }.
\end{equation}

\subsection{Partition function}

We start with considering the partition function of a circular polymer model with an effective Hamiltonian (\ref{Hdis}).
Performing an expansion in coupling constants  $b_0$ and $w_0$ in the exponent and keeping terms up to the 1st order one has:
\begin{eqnarray}
Z_{{\rm ring}}(S)=Z^0(S)-b_0 Z^1_{b_0}(S) + w_0 Z^1_{w_0}(S). \label{Zex}
\end{eqnarray}
Here, $Z^0(S)$ is the partition function of an idealized ``unperturbed'' Gaussian model without any interactions
between monomers:
\begin{eqnarray}
&&Z^0_{{\rm ring}}(S){=}\frac{1}{Z^0_{{\rm o}}}\! \int D\vec{r}\, \delta (\vec{r}(S)-\vec{r}(0))\, {\rm e}^{-\frac{1}{2}\int_0^S ds\left(\frac{d\vec{r}(s)}{ds}\right)^2} \phantom{555}\label{Z}
\end{eqnarray}
normalized in such a way that the partition function of an open Gaussian chain is unity, here $Z^0_{{\rm o}}\equiv
\int D\vec{r}\,{\rm e}^{  -\frac{1}{2}\int_0^S ds\left(\frac{d\vec{r}(s)}{ds}\right)^2 } $.

Exploiting the Fourier-transform of the  $\delta$-function
\begin{equation}
\delta (\vec{r}(S)-\vec{r}(0)) =\frac{1}{(2\pi)^{d}} \int {\rm d}\vec{q}\, {\rm e}^{\left(-i\vec{q}(\vec{r}(S)-\vec{r}(0)\right)} \label{d}
\end{equation}
and rewriting: $\vec{r}(S)-\vec{r}(0)=\int_0^S ds\left(\frac{d\vec{r}(s)}{ds}\right)$, one has:
\begin{eqnarray}
&&Z^0_{{\rm ring}}(S)=\frac{1}{Z^0_{{\rm o}}}\int D\vec{r}\, {\rm e} ^{-\frac{1}{2}\int_0^S ds\left(\frac{d\vec{r}(s)}{ds}-i\vec{q}\right)^2 }\times\nonumber\\
&& \times \frac{1}{(2\pi)^{d}}\int {\rm d}\vec{q}\, {\rm e}^{-\frac{q^2S}{2}}.
\end{eqnarray}
After performing the Poisson integration, one easily obtains the partition function of Gaussian ring polymer \cite{Duplantier94}:
\begin{equation}
Z^0(S)= (2\pi S)^{-d/2}.
\end{equation}

The contribution $Z^1_{b_0}(S)$ into the perturbation theory expansion (\ref{Zex}) is given by:
\begin{eqnarray}
&&Z^1_{b_0}(S)=\frac{1}{Z^0_{{\rm o}}}\int_0^S\!\!{\rm d}s' \int_0^{s'}\!\! {\rm  d}s'' \int D\vec{r}\, {\rm e} ^{-\frac{1}{2}  \int_0^S ds\left(\frac{d\vec{r}(s)}{ds}\right)^2 }
\times \nonumber\\
&&\times\frac{1}{(2\pi)^{2d}} \int {\rm d}\vec{q}\, {\rm e}^{-i\vec{q}(\vec{r}(S)-\vec{r}(0))}
 \int {\rm d}\vec{k}\, {\rm e}^{-i\vec{k}(\vec{r}(s')-\vec{r}(s''))} = \nonumber\\
 && =\frac{1}{(2\pi)^{2d}}\int_0^S\!\!{\rm d}s' \int_0^{s'}\!\! {\rm  d}s'' \int d \vec{k} \int d \vec{q} \,
 {\rm e}^{-\frac{(\vec{k}+\vec{q})^2}{2}(s'-s'')}\times \nonumber\\
 && \times \,\,{\rm e}^{-\frac{\vec{q}^2}{2}(S-s'+s'')} =\nonumber\\
&&=(2 \pi)^{-d} \int_0^S\!\!{\rm d}s' \int_0^{s'}\!\! {\rm  d}s'' (s'-s'')^{-\frac{d}{2}}(S-s'+s'')^{-\frac{d}{2}}\nonumber\\
&&=(2 \pi)^{-d} S^{2-d} B(1-{d}/{2},2-{d}/{2}), \label{scheme}
\end{eqnarray}
here $B(1-d/2,2-d/2)$ is Euler beta function.

Taking into account that the Fourier transform of the correlation function (\ref{avv0}) in the limit of large $\vec{r}$ is 
  {(see Appendix A)}:
\begin{eqnarray}
&&|\vec{r}(s')-\vec{r}(s'')|^{-a} \simeq \int d\vec{k} \,|\vec{k}|^{a-d}  {\rm e}^ {-i\vec{q}(\vec{r}(s')-\vec{r}(s''))},\nonumber
\end{eqnarray}
 the contribution $Z^1_{w_0}(S)$ can be easily evaluated according to the scheme (\ref{scheme}):
  {
 \begin{eqnarray}
&&Z^1_{w_0}(S)=\frac{1}{(2\pi)^{d}}\int d \vec{k} |k|^{a-d} \int_0^S\!\!{\rm d}s' \int_0^{s'}\!\! {\rm  d}s'' 
 {\rm e}^{\frac{|k|^2}{2S}(s'-s'')} \nonumber\\
 && \times\,\, {\rm e}^{-\frac{|k|^2}{2}(s'-s'')} \equiv \int {\rm d} \vec{k} |k|^{a-d} f(|\vec{k}|). \label{s1}
\end{eqnarray}
Passing to $d$-dimensional spherical coordinate system, integration over $\vec{k}$ can be presented as:
\begin{eqnarray}
\int {\rm d} \vec{k}  \rightarrow  \Omega_{d} \int_0^{\infty}{\rm d } k |k|^{d-1},
\end{eqnarray}
where $\Omega_d$ denotes integration over angular variables. 
Thus, in the above expression we have:
\begin{eqnarray}
&& \int {\rm d} \vec{k}\, |k|^{a-d} f(|\vec{k}|) =  \Omega_{d} \int_0^{\infty} {\rm d} k |k|^{a-1} f(|\vec{k}|). \label{s2}
\end{eqnarray}
Due to the fact that $f(|\vec{k}|)$ depends only on module of $\vec{k}$, we immediately conclude that, except of angular factor $\Omega_d$ which can be adsorbed into
redefinition of coupling constant, the integrals over $\vec{k}$ can be treated as $a$-dimensional. }

 The final expression for a partition function of a ring polymer then reads:
\begin{eqnarray}
&&Z_{{\rm ring}}(S)=(2 \pi S)^{-{d}/{2}} \left(1-z_{b_0}B(1-{d}/{2},2-{d}/{2})\right.\nonumber \\
&&\left.+z_{w_0}B(1-{a}/{2},2-{a}/{2})\right),
\end{eqnarray}
here, the dimensionless couplings are introduced:
 \begin{equation}
z_{b_0}=b_0(2\pi)^{-d/2}S^{2-d/2}, z_{w_0}=w_0(2\pi)^{-a/2}S^{2-a/2}.
\end{equation}
Similarly  for the partition function of an open linear chain we have:
 \begin{equation}
Z_{{\rm chain}}(S)=1-\frac{z_{b_0}}{(1-{d}/{2})(2-{d}/{2})}
+\frac{z_{w_0}}{(1-{a}/{2})(2-{a}/{2})}.
\end{equation}

\subsection{Gyration radius and $g$-ratio}

To evaluate the expression for
gyration radius as given by (\ref{rg}), we start by rewriting:
\begin{eqnarray}
(\vec{r}(s_2)-\vec{r}(s_1))^2=-{2d}\frac{{\rm d}}{{\rm d} k^2}
 {\rm e}^{-i\vec{k}(\vec{r}(s_2)-\vec{r}(s_1))}|_{k=0}. \label{defin}
\end{eqnarray}
In the ``unperturbed'' Gaussian approximation one has in the case of closed ring polymer:
\begin{eqnarray}
&&\langle  {\rm e}^{-i\vec{k}(\vec{r}(s_2)-\vec{r}(s_1))} \rangle^0 =  {\rm e}^{-\frac{{k}^2}{2}\frac{(s_2-s_1)(S-s_2+s_1)}{S}},
\nonumber
\end{eqnarray}
and thus:
\begin{eqnarray}
&&\langle R_{g\,{\rm ring}}^2\rangle^0=\frac{d}{S^2}\int_0^S ds_2 \int_0^{s_2} ds_1 \left(s_2-s_1-\right.\nonumber\\
&&\left.-(s_2-s_1)^2/S\right)=\frac{Sd}{12}.
\end{eqnarray}

\begin{figure}
\includegraphics[width=85mm]{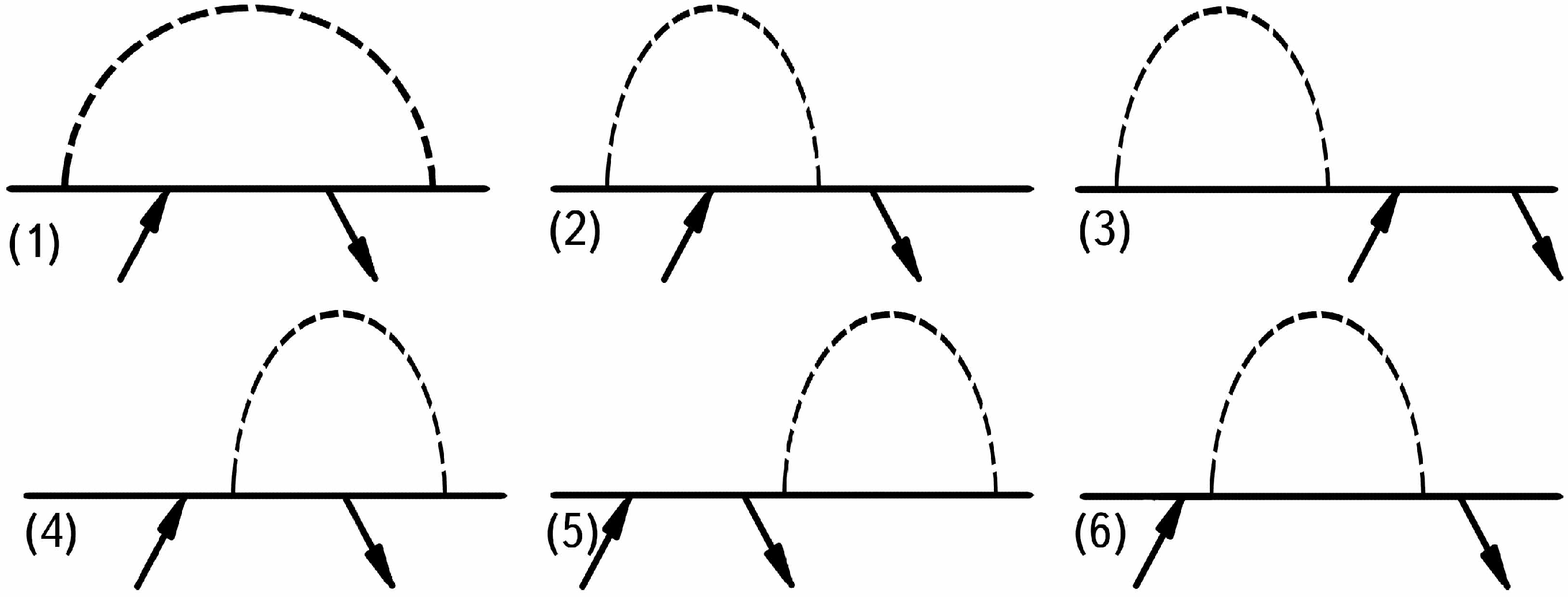}
\caption{Diagrammatic presentation of contributions into the gyration radius of ring polymer up to the first order in coupling constants.
 Dashed lines denote the monomer-monomer interactions. Each diagram appears twice: once with excluded volume interaction governed by coupling $z_{b_0}$ and once with disorder interaction $z_{w_0}$. By arrows we denote the restriction points $s_1$ and $s_2$ bearing incoming and outcoming wave vector $k$.}
\label{fig:1}
\end{figure}

Performing the perturbation theory expansion in coupling constants and keeping terms up to the 1st order one may thus write:
\begin{equation}
\langle R_{g\,{\rm ring}}^2\rangle = \frac{Sd}{12} - z_{b_0}\Omega^1_{z_{b_0}} + z_{w_0}\Omega^1_{z_{w_0}}.
\end{equation}
In what follows we will use the diagrammatic presentation of perturbation theory series, and
exploit the same diagrams for both chains and rings (see Fig. \ref{fig:1}).
Thus, $\Omega^1_{z_{b_0}}$ and $\Omega^1_{z_{w_0}}$ correspond
to contributions, presented by diagrams on Fig. \ref{fig:1} with pairwise interactions between monomers
governed by coupling constants $z_{b_0}$ and $z_{w_0}$ respectively and have general form (see Appendix B for details):
\begin{eqnarray}
\Omega^1_{z_{b_0}}=\sum_i \alpha_i B(\beta_i-d/2,\gamma_i-d/2), \nonumber\\
\Omega^1_{z_{w_0}}=\sum_i \alpha'_i B(\beta'_i-a/2,\gamma'_i-a/2),
\end{eqnarray}
where the Greek symbols denote  rational  and natural numbers.

Proceeding with the double $\varepsilon=4-d$, $\delta=4-a$ expansions of above expressions, we obtain:
\begin{equation}
\langle R_{g\,{\rm ring}}^2\rangle = \frac{Sd}{12} \left(1+\frac{2z_{b_0}}{\varepsilon}-\frac{2z_{w_0}}{\delta}\right).\label{rrgfinl}
\end{equation}
Applying the same scheme as described above, for the radius of gyration of (open) linear chain one has:
\begin{equation}
\langle R_{g\,{\rm chain}}^2\rangle = \frac{Sd}{6} \left(1+\frac{2z_{b_0}}{\varepsilon}-\frac{13z_{b_0}}{12}-\frac{2z_{w_0}}{\delta}+\frac{13z_{w_0}}{12}\right)
\end{equation}
Thus, we obtain the estimate of the size ratio (\ref{ratioR}) up to the first order of perturbation theory expansion:
\begin{equation}
g \equiv \frac{\langle R_{g\,{\rm ring}}^2\rangle}{\langle R_{g\,{\rm chain}}^2\rangle}= \frac{1}{2} \left(1+\frac{13z_{b_0}}{12}-\frac{13z_{w_0}}{12}\right).
\label{vidn}
\end{equation}

The  universal properties of linear and ring polymers are known to be governed by the same
fixed points values (\ref{fip}) within the renormalization group scheme \cite{Prentis82}.
Thus we make use of results for fixed point values found previously for the linear polymer chains
in long-range correlated disorder \cite{Blavatska}.  There are three distinct fixed points governing the properties of
macromolecule in various regions of parameters $d$ and $a$:
\begin{eqnarray}
&& {\mbox {Gaussian}}:\,\,\, z^*_{b_0}=0,  z^*_{w_0}=0, \label{fp1}\\
&& { \mbox {Pure}}:\,\,\,\,\,\,\,\,\,\,\,\,\,\, z^*_{b_0}=\frac{\varepsilon}{8}, z^*_{w_0}=0, \\
&& {\mbox  {mixed LR}}: z^*_{b_0}=\frac{\delta^2}{4(\varepsilon-\delta)},  z^*_{w_0}=\frac{\delta(\varepsilon-2\delta)}{4(\delta-\varepsilon)}. \label{fp3}
\end{eqnarray}
Evaluating Eq. (\ref{vidn}) in these three cases, we obtain:
\begin{eqnarray}
&&g^{{\rm Gauss}}  = \frac{1}{2},\label{resultgaus}\\
&&g^{{\rm pure}} = \frac{1}{2} \left(1+\frac{13}{96}\varepsilon\right),\label{resultpure}\\
&&g^{{\rm LR}} = \frac{1}{2} \left(1+\frac{13}{48}\delta\right).\label{resultrg}
\end{eqnarray}
With  $g^{{\rm pure}}$ we recover the result found previously in Ref. \cite{Prentis82} for the
 polymers in pure solution, whereas the
last expression gives the value of size ratio in the solution in presence of long-range correlated
structural obstacles.  { Whereas the  fixed points coordinates (\ref{fp3}) depend on  both of the global parameters $d$ and $a$, the resulting size ratio (\ref{resultrg}) in the one-loop approximation depends only on $a$.
   Note, however, that disorder characterized by some fixed value of parameter $a$,  would correspond  to
 different physical realizations depending on the space dimension. Really, remembering that correlation function in the form (5) refers to extended defects of
(fractal) dimension $d_f=d-a$,  the same value $a=1$ corresponds to planar impurities in $d=3$ and linear defects in $d=2$, respectively. Based on this consideration, one may say, that relations like (\ref{resultrg}) indirectly
imply dependence on $d$.}

To find the quantitative estimate for the size ratio $g$ in pure solution in $d=3$, we evaluate the expression (\ref{resultpure}) at $\varepsilon=1$ and obtain $g^{{\rm pure}}\simeq 0.57$. One may easily convince oneself, that presence of long-range correlated disorder with any $a<d$ leads to an
increase of this value, as given by  (\ref{resultrg}).
Moreover,  this ratio grows with an increasing strength of disorder (decreasing of parameter $a$), and thus the distinction between
the size measure of a ring and an open linear polymers of the same molecular weight is smaller in disordered environment as compared with the pure solution.
  {From physical point of view, we can interpret this as follows. The presence of obstacles in environment is expected to produce an effective entropic attraction between monomers of macromolecules (see (\ref{Hdis})). However, the case of long-range correlated disorder corresponds to complex (fractal) defects extended throughout the system. The polymer macromolecule is forced to avoid these extended regions of space, which results in its elongation
and an increase of the shape anisotropy of a typical polymer conformation in such disordered environment.}

\subsection{Spanning radius and $p$-ratio}

Another interesting characteristic of a size measure of a circular polymer is the spanning radius $\langle R_{1/2}^2 \rangle$.
To evaluate the expression for $\langle R_{1/2}^2 \rangle$
 as given by (\ref{re}), we start by rewriting:
\begin{eqnarray}
(\vec{r}(S/2)-\vec{r}(0))^2=-{2d}\frac{{\rm d}}{{\rm d} k^2} {\rm e}^{-i\vec{k}(\vec{r}(S/2)-\vec{r}(0))}|_{k=0}. \label{Rmid}
\end{eqnarray}
In the ``unperturbed'' Gaussian approximation we have:
\begin{eqnarray}
&&\langle  {\rm e}^{-i\vec{k}(\vec{r}(S/2)-\vec{r}(0))} \rangle^0 =  {\rm e}^{-\frac{{k}^2S}{8}},
\nonumber
\end{eqnarray}
and thus:
\begin{eqnarray}
&&\langle R_{1/2\,{\rm ring}}^2\rangle^0=\frac{Sd}{4}.
\end{eqnarray}

\begin{figure}
\includegraphics[width=85mm]{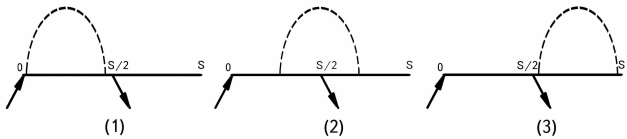}
\caption{Diagrammatic presentation of contributions into the spanning radius of ring polymer up to the first order in coupling constants.
 Notations are the same as in Fig. 1.}
\label{fig:6}
\end{figure}

Again, we use diagrammatic presentation of  contributions into the spanning radius, produced by perturbation theory expansion in coupling constants
 (see Fig. \ref{fig:6}).

 %As an example of calculation we will consider third diagram from this figure:
%\begin{eqnarray}
%&&\int d\vec{q} \int d\vec{p}\int_{S/2}^S ds' \int_0^{s'} ds'' %exp\left(-\frac{(\vec{q}+\vec{k})^2}{2}\frac{S}{2}\right.\\ \nonumber %&&\left.-\frac{(\vec{q}+\vec{p})^2}{2}\left(s'+s''\right)-\frac{\vec{q}^2}{2}\left(\frac{S}{2}-s'+s''\right)\right)
%\end{eqnarray}
%Integrating over wave vectors and using a definition (\ref{RL}) will lead to:
%\begin{eqnarray}
%&& \int_{S/2}^S ds' \int_0^{s'} ds''(s'-s'')^{-d/2}(S-s'+s'')^{-d/2}\\ \nonumber
%&&\left(\frac{S}{4} -\frac{(s'+s'')^2}{4(s'-s'')(S-s'+s'')}\right)
%\end{eqnarray}
%Presenting a dimensionless variables and making a change $t=s'-s''$ and $s=s'-1/2$ one will came to:
%\begin{eqnarray}
%&&\frac{S^{3-d}}{4} \int_0^{1/2} ds \int_0^{s} dt(t)^{-d/2}(1-t)^{-d/2}\\ \nonumber
%&&\left(1 -\frac{t}{4(1-t)}\right)
%\end{eqnarray}
%Due to the fact that only limit of integration depends on $s$ one can came to:
%\begin{eqnarray}
%&&\frac{S^{3-d}}{4} \int_0^{1/2} dt \left(\frac{1}{2}-t\right)t^{-d/2}(1-t)^{-d/2}\\ \nonumber
%&&\left(1 -\frac{t}{(1-t)}\right)
%\end{eqnarray}
%And this integral is nothing else but some of four incomplete Beta functions:
%\begin{eqnarray}
%&&\frac{S^{3-d}}{4} \left(\frac{1}{2}B_{1/2}(1-d/2,1-d/2) - B_{1/2}(2-d/2,1-d/2)\right.\\ \nonumber
%&&\left.\frac{1}{2}B_{1/2}(2-d/2,-d/2) - B_{1/2}(3-d/2,-d/2)\right) \nonumber
%\end{eqnarray}

Applying the same scheme for diagram calculation, as described in previous subsection and Appendix B,
we found
\begin{equation}
\langle R_{1/2}^2 \rangle = \frac{Sd}{4} \left(1+\frac{z_{b_0}}{2}+\frac{2z_{b_0}}{\varepsilon}-\frac{z_{w_0}}{2}-\frac{2z_{w_0}}{\delta}\right).
\end{equation}
Recalling expression for the gyration radius of ring polymer (\ref{rrgfinl}):
\begin{equation}
p\equiv \frac{\langle R_{1/2\,{\rm ring}}^2\rangle}{\langle R_{g\,{\rm ring}}^2\rangle} =  3 \left(1+\frac{z_{b_0}}{2}-\frac{z_{w_0}}{2}\right).
\end{equation}
Evaluating this ratio  at fixed points (\ref{fp1})-(\ref{fp3}),  governing the properties of
macromolecule in various regions of parameters $d$ and $a$, we finally have:
\begin{eqnarray}
&&p^{{\rm Gauss}}  = 3,\label{resultgaus1/2}\\
&&p^{{\rm pure}} = 3 \left(1+\frac{\varepsilon}{16}\right),\label{resultpure1/2}\\
&&p^{{\rm LR}} = 3 \left(1+\frac{\delta}{8}\right).\label{resultlr1/2}
\end{eqnarray}
To find the qualitative estimate for the size ratio $p$ in pure solution in $d=3$, we evaluate the expression (\ref{resultpure1/2})
at $\varepsilon=1$ and obtain $p^{{\rm pure}}\simeq 3.18$, which is in nice agreement
with result of computer simulations, found previously in Ref. \cite{Jagodzinski92}: $p^{{\rm pure}}=3.217\pm0.020$. Again, the presence of long-range correlated disorder with any $a<d$ leads to an
increase of this value, as given by  (\ref{resultlr1/2}): this ratio grows with an increasing strength of disorder
(decreasing of parameter $a$).

\section{Conclusions}

In the present paper we analyze the statistical properties of flexible polymers in a form of closed rings   in solutions in presence of structural
obstacles (impurities). One  encounters such situations when considering polymers in gels, colloidal solutions
 or in the cellular environment. We consider a special case of so-called long-range correlated disorder, assuming the defects to be correlated on large distances $r$
according to a power law with a pair correlation function $
g(r)\sim r^{-a} $ \cite{Weinrib83}. For $a<d$, such a correlation function describes defects extended in space, which form complex structures of
(fractal) dimension $d_f=d-a$, such that
$a=d-2$ $(d-1)$ correspond respectively to the impurities in form of lines (planes), randomly distributed in  space, whereas non-integer values of $a$ refer to defects of fractal structure.

Applying a  direct polymer renormalization scheme, we study  the universal size and shape characteristics of macromolecules,
such as the size ratios (\ref{ratioR}) and  (\ref{ratioRpiv}).
Our results reveal an essential influence of disorder on the spatial extension and anisotropy of typical circular polymer conformation.
In particular, the presence of long-range correlated disorder with any $a<d$ leads to an
increase  of the size ratio of a ring and an open linear polymers of the same molecular weight as given by  (\ref{resultrg}):
this value grows with an increasing strength of disorder (decreasing of parameter $a$), and thus the distinction between
the size measure of circular and open chains is smaller in disordered environment as compared with the pure solution.
 From physical point of view, this can interpreted  as follows.
The case of long-range correlated disorder corresponds to complex (fractal) defects extended throughout
the system. The polymer macromolecule is forced to avoid these extended regions of space, which results in its elongation
and an increase of the shape anisotropy of a typical polymer conformation in such disordered environment.

\section*{Acknowledgements}

This work was supported in part by the
FP7 EU IRSES projects N269139  ``Dynamics and Cooperative Phenomena in Complex
Physical and Biological Media'' and N295302 ``Statistical Physics in Diverse Realizations''.

  {
\section*{Appendix A}
Here, we evaluate the Fourier transformation of correlation function (\ref{avv0}) in the form $|r|^{-a}$:
\begin{eqnarray}
\xi(\vec{k}) = \int {\rm d}\vec{r}\, |\vec{r}|^{-a} \exp( i \vec{k}\vec{r}).\nonumber
\end{eqnarray}
Passing to $d$-dimensional spherical coordinate system and performing integration over angular variables one has:
\begin{eqnarray}
&&\xi(\vec{k}) = \frac{2\pi ^{\frac{d-1}{2}}}{\Gamma(\frac{d-1}{2})}\int_0^{\pi} {\rm d }\theta \,(\sin \theta)^{d-2}\times\nonumber  \\
&&\int_0^{\infty} {\rm d}r\, |\vec{r}|^{d-1-a} \exp( i k r \cos \theta). \label{fur1} 
\end{eqnarray}
Introducing the variable $x\equiv kr$, the last integration  in (\ref{fur1}) can be
rewritten as:
\begin{eqnarray}
k^{a-d}\int_0^{\infty} {\rm d}x\, x^{d-1-a} \exp( i x \cos \theta). \nonumber 
\end{eqnarray}
Making use of relation (3.915(5)) in Ref. \cite{Gradstein}, we have:
\begin{eqnarray}
&&\int_0^{\pi} {\rm d }\theta \,\exp( i x \cos \theta)(\sin \theta)^{d-2} =\nonumber\\
&&= \sqrt{\pi}\left(\frac{2}{x}\right)^{\frac{d-2}{2}}\Gamma\left(\frac{d-1}{2}\right)J_{\frac{d-2}{2}}(x),
 \nonumber 
\end{eqnarray}
 here $J_{\frac{d-2}{2}}(x)$ is Bessel function.
 Finally, we use the relation (6.561(14)) in Ref. \cite{Gradstein} to evaluate:
\begin{eqnarray}
\int_0^{\infty} {\rm d}x\, x^{\frac{d-2a}{2}} J_{\frac{d-2}{2}}(x) = 2^{\frac{d-2a}{2}}\frac{\Gamma(\frac{d-a}{2})}{\Gamma(\frac{a}{2})}.\nonumber 
\end{eqnarray}
 This will lead to the following form of Fourier transform:
\begin{eqnarray}
&&\xi(\vec{k}) = k^{a-d}\pi^{\frac{d}{2}}2^{d-a}\frac{\Gamma(\frac{d-a}{2})}{\Gamma(\frac{a}{2})}.
\end{eqnarray}
}

\section*{Appendix B}

\begin{figure}[t!]
\begin{center}
\vspace*{-1.5cm}
\includegraphics[width=80mm]{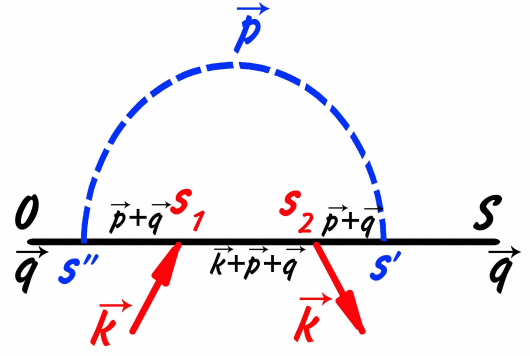}
\end{center}
\vspace*{-2cm}
\caption{Example of diagrammatic contribution into the gyration radius of ring polymer. }
\label{fig:5}
\end{figure}

Here, we evaluate an analytic expression corresponding to the diagram (1) on Fig. 1 which produces contributions
into the gyration radius of ring polymer (shown in more details on Fig. \ref{fig:5}).
The solid line on a diagram is a schematic presentation of a polymer 	of  length $S$,
  {dashed line denotes the long-range
interaction governed by  coupling
$z_{w_0}$} between points $s'$ and $s''$  (interaction points), and $s_1$ and $s_2$ are so-called restriction points. According to
the general rules of diagram calculations \cite{desCloiseaux}, each segment between any two points $s_a$ and $s_b$
is oriented and bears a wave vector $\vec{q}_{ab}$ given by a sum of incoming and outcoming  wave vectors injected
at interaction points, restriction points and end points.
At these points, the flow of wave vectors is
conserved.
A factor $\exp\left(-\frac{\vec{q}_{ab}^2}{2}(s_b-s_a)\right)$ is accosiated with each segment, and integration is to be made
over all independent segment areas and over wave vectors injected at the end points and interaction points.
The diagram shown on Fig. \ref{fig:5} is than associated with an expression:
\begin{eqnarray}
&&\frac{1}{S^{2}(2\pi)^{2d}}\int_0^S\!\! {\rm d }s'\int_{0}^{s'}\!\!{\rm d }s''\!\! \int d\vec{q}\! \int d\vec{p}  { |p|^{a-d}}
\, {\rm e}^ {-\frac{\vec{q}^2 }{2}(S-s'+s'')} 
\nonumber\\
&&\times\int_{s''}^{s'}\!\!{\rm d }s_2\!\!\int_{s''}^{s_2}\!\! {\rm d }s_1\, {\rm e}^{ -\frac{(\vec{q}+\vec{p})^2}{2}(s'-s_2+s_1-s'')-\frac{(\vec{k}+\vec{q}+\vec{p})^2}{2}(s_2-s_1)}. \nonumber
\end{eqnarray}
Performing the Gaussian integration over the wave vectors $\vec{p}$ and $\vec{q}$, taking into account the pecularities of 
calculation the  contributions of long-range correlated disorder 
(\ref{s1}) -  (\ref{s2}), after taking derivative over $k$ according to (\ref{defin}) we have:
\begin{eqnarray}
&& \int_0^S \!\!{\rm d }s'\!\!\int_{0}^{s'}\!\!{\rm d }s''\,(s'-s'')^{-a/2}(S-s'+s'')^{-a/2}\times \nonumber\\
&& \times\int_{s''}^{s'}ds_2\int_{s''}^{s_2}ds_1\left(s_2-s_1-\frac{(s_2-s_1)^2}{s'-s''}\right), \nonumber
\end{eqnarray}
note that prefactor $S^{-2}(2\pi)^{\frac{a}{2}-\frac{3d}{2}}$ is omitthed in previous expression.
Integrating this expression over $s_1$ and $s_2$ and passing to dimensionless
variables $h'=s'/S$, $h=h'-s''/S$ we obtain:
\begin{eqnarray}
&&I_1=\frac{1}{12}\int_0^1 {\rm d } h'\!\int_{0}^{h'}\!{\rm d }h\, h^{3-a/2}(1-h)^{-a/2}=\nonumber \\
&&=\frac{1}{12}\int_0^1 {\rm d }\, h' B_{h'}(4-a/2,1-a/2), \nonumber
\end{eqnarray}
where the definition  of an incomplete Euler beta function $B_{s}(a,b)= \int_0^s x^{a-1}(1-x)^{b-1} {\rm d} x$ is used.
In dealing with integration of this type, we make use of relation \cite{Prudnikovbook}:
\begin{eqnarray}
&&\int {\rm d}s\, s^{\lambda} B_s(a,b)= \frac{s^{\lambda+1}}{\lambda+1} B_s(a,b)-\nonumber\\
&&-\frac{1}{\lambda+1} B_s(a+\lambda+1,b). \label{B}
\end{eqnarray}
Thus,  the result for this diagram reads:
\begin{eqnarray}
&&I_1=\frac{1}{12}\Big[B(4-a/2, 1-a/2)-B(5-a/2, 1-a/2) \Big]. \nonumber
\end{eqnarray}
 To proceed with $\delta$-expansion of such expressions, we recall that $B(a,b)=\Gamma(a)\Gamma(b)/\Gamma(a+b)$ and exploit the expansion of Euler gamma function:  $\Gamma(1+x)\sim 1-cx$ with Euler constant $c\simeq 0.5772$. Applying this scheme to the expression above, we obtain:
    \begin{eqnarray}
I_1=\frac{1}{12}\frac{(2+\delta)\Gamma^2(1+\delta/2)}{\delta \Gamma(1+\delta) (1+\delta)}\simeq\frac{1}{6\delta}-\frac{1}{12}+O(\delta).
\nonumber
\end{eqnarray}
The expressions corresponding to other diagrams $(2)$-$(6)$ on Fig. \ref{fig:1} with the long-range
interaction governed by coupling $z_{w_0}$, evaluated within the same scheme, read:
\begin{eqnarray}
&&I_2=I_4=\frac{1}{12}\Big[2B(1-a/2, 2-a/2)- \nonumber\\
&&-B(1-a/2, 5-a/2)-5B(1-a/2, 3-a/2) +\nonumber\\
&&+4B(1-a/2, 4-a/2)-B(-a/2, 2-a/2)- \nonumber\\
&&-B(-a/2, 6-a/2)+4B(-a/2, 5-a/2)- \nonumber\\
&&-6B(-a/2, 4-a/2)+4B(-a/2, 3-a/2) \Big]= \nonumber\\
&&\simeq\frac{1}{6\delta}+O(\delta);\\
&&I_3=I_5=\frac{1}{240}\Big[ 10B(1-a/2, 1-a/2)- \nonumber\\
&&-40B(1-a/2, 4-a/2)+60B(1-a/2, 3-a/2)- \nonumber\\
&&-40B(1-a/2, 2-a/2)-4B(-a/2, 1-a/2) +\nonumber\\
&&+4B(-a/2, 6-a/2)-20B(-a/2, 5-a/2)+ \nonumber\\
&&+40B(-a/2, 4-a/2)-40B(-a/2, 3-a/2)+ \nonumber\\
&&+20B(-a/2, 2-a/2)+10B(1-a/2, 5-a/2) \Big]= \nonumber\\
&&\simeq-\frac{1}{10\delta}-\frac{1}{40}+O(\delta);\\
&&I_6=\frac{1}{60}\Big[5B(1-a/2, 1-a/2) + \nonumber\\
&&+5B(1-a/2, 5-a/2)-20B(1-a/2, 2-a/2)+ \nonumber\\
&&-20B(1-a/2, 4-a/2)+30B(1-a/2, 3-a/2) -\nonumber\\
&&-3B(-a/2, 1-a/2)+3B(-a/2, 6-a/2)- \nonumber\\
&&-15B(-a/2, 5-a/2)+15B(-a/2, 2-a/2) -\nonumber\\
&&-30B(-a/2, 3-a/2)+30B(-a/2, 4-a/2) \Big]= \nonumber\\
&&\simeq-\frac{2}{15\delta}-\frac{1}{30}+O(\delta).
\end{eqnarray}
The expressions corresponding to the diagrams with monomer-monomer interactions governed by excluded volume coupling
$z_{b_0}$ are  easily obtained
when replacing correlation parameter $a$  by space dimension $d$ in above expressions for $I_1$-$I_6$.

\end{document}